\documentclass[runningheads]{llncs}
\usepackage{xcolor}
\usepackage[misc]{ifsym}
\usepackage[T1]{fontenc}
\usepackage{graphicx}
\begin{document}
\title{FingerNet: EEG Decoding of A Fine Motor Imagery with Finger-tapping Task Based on \\ A Deep Neural Network}
%
%
\author{Young-Min Go\inst{1}\orcidID{0009-0009-2303-918X} \and
Seong-Hyun Yu\inst{1}\orcidID{0000-0003-3455-9753} \and
Hyeong-Yeong Park\inst{1}\orcidID{0000-0001-6943-4171} \and
Minji Lee\inst{2}\orcidID{0000-0003-4261-875X} \and
Ji-Hoon Jeong*\inst{1(\resizebox{0.7em}{0.4em}{\textrm{\Letter}})}\orcidID{0000-0001-6940-2700}}
%

%
\authorrunning{Young-Min et al.}
\titlerunning{FingerNet: Decoding of A Fine Motor Imagery Paradigm from EEG Signals}

\institute{Dept. Computer Science, Chungbuk National University, Chungbuk 28644, Republic of Korea\\
\email{\{ym.go,sh.yu,hyeong.y.park,jh.jeong\}@chungbuk.ac.kr}\\ \and
Dept. Biomedical Software Engineering, The Catholic University of Korea, Gyeonggi 14662, Republic of Korea\\
\email{minjilee@catholic.ac.kr}
}
\maketitle              
\begin{abstract}
Brain-computer interface (BCI) technology facilitates communication between the human brain and computers, primarily utilizing electroencephalography (EEG) signals to discern human intentions. Although EEG-based BCI systems have been developed for paralysis individuals, ongoing studies explore systems for speech imagery and motor imagery (MI). This study introduces FingerNet, a specialized network for fine MI classification, departing from conventional gross MI studies. The proposed FingerNet could extract spatial and temporal features from EEG signals, improving classification accuracy within the same hand. The experimental results demonstrated that performance showed significantly higher accuracy in classifying five finger-tapping tasks, encompassing thumb, index, middle, ring, and little finger movements. FingerNet demonstrated dominant performance compared to the conventional baseline models, EEGNet and DeepConvNet. The average accuracy for FingerNet was 0.3049, whereas EEGNet and DeepConvNet exhibited lower accuracies of 0.2196 and 0.2533, respectively. Statistical validation also demonstrates the predominance of FingerNet over baseline networks. For biased predictions, particularly for thumb and index classes, we led to the implementation of weighted cross-entropy and also adapted the weighted cross-entropy, a method conventionally employed to mitigate class imbalance. The proposed FingerNet involves optimizing network structure, improving performance, and exploring applications beyond fine MI. Moreover, the weighted Cross Entropy approach employed to address such biased predictions appears to have broader applicability and relevance across various domains involving multi-class classification tasks. We believe that effective execution of motor imagery can be achieved not only for fine MI, but also for local muscle MI. 

\end{abstract}
\section{Introduction}

Brain-computer interface (BCI) is a promising technology of communication between human brain and computers. BCI technologies generally employ electroencephalography (EEG) signals to identify human intentions. Noninvasive Brain-Machine Interfaces (BMIs) aim to encode human cognitive processes, extracted from brain activity, into control signals, thereby facilitating external applications without the requirement for invasive brain implantation surgery. EEG-based BCI systems have been developed for individuals with paralysis or disabilities \cite{ref_article1,add_1_1,add_1_2,add_1_3,add_1_4}. Studies are in progress for various systems, including speech imagery to classify intended words through imagined speech and motor imagery (MI) to classify desired movements through imagined actions \cite{ref_article2,ref_article3,ref_article4,add_4_1}. The development is progressing towards identifying the intentions of patients experiencing various discomforts, aiming to assist in rehabilitation or application. Advancements in these technologies offer benefits not only for the rehabilitation and application of patients with paralysis but also for enhancing the capabilities of healthy individuals \cite{ref_article5,ref_article6,add_6_1,add_6_2,add_6_3,add_6_4}.

MI, one of the primary research fields of BCI, is a domain focused on identifying and classifying the intended movements of users. In existing MI research, studies are mainly focused on gross motor. \cite{new_1,new_2,new_3,new_4,new_5} Among them, various experiments using the upper limb have been conducted extensively, such as arm flexion/extension, hand grasping, and shoulder movements \cite{ref_article7,ref_article8,add_8_1,add_8_2}. Recent research on MI-based BCI has been demonstrated the applicability through connectivity with external devices. We endeavored to integrate MI-based BCI into the domain of fine motor imagery. In our study, we explored the domain of fine MI, departing from the conventional focus on gross MI.

In this paper, we propose a FingerNet for the classification of finger-tapping MI. Some related studies have focused on simple classifications, such as distinguishing between the right and left hand. However, classifications within the same hand have not been conducted extensively. We expanded the range of conventional experiments to address the classification of individual finger classes within a right hand. In previous research, performance was assessed using machine learning techniques and EEGNet, but achieving high accuracy proved to be a challenge. We predicted difficulties in classifying closely placed classes, because EEGNet tends to predominantly utilize spatial features for classification \cite{ref_article10,add_10_1,add_10_2}. Therefore, we designed a network that could complement these weaknesses of EEGNet. We refer to concepts from DeepConvNet, improving the EEGNet model by adding convolution and pooling layers to extract additional temporal features \cite{ref_article11,add_11_1,add_11_2,add_11_3}. Additionally, we observed bias predictions occurring for specific classes in the experimental results. To address these issues, we applied a weighted cross-entropy loss function. Weighted cross-entropy has been conventionally used in situations where performance is compromised due to class imbalance \cite{ref_article12,add_12_1,add_12_2}. However, we attempted to address biased predictions of such a model by assigning weights to the loss values. We adjusted the weights for each class heuristically to find optimal values, effectively resolving bias predictions. 

\section{Materials and Methods}

\subsection{Experimental Protocols}

\begin{figure}[t!] 
\centering
\includegraphics[width=0.9\textwidth]{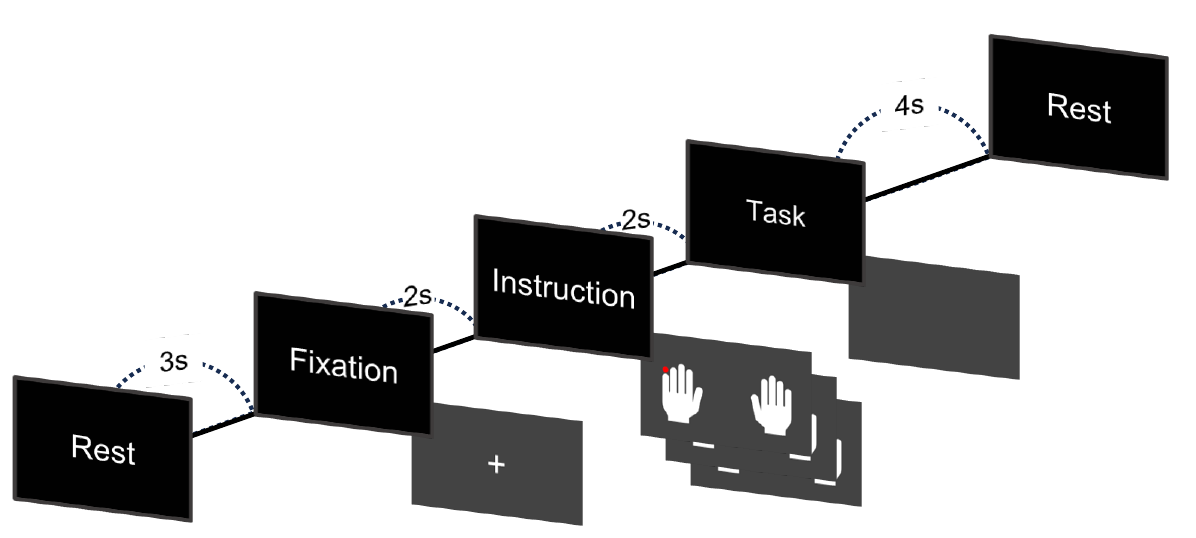}
\caption{Experimental protocol comprises the phases of rest (3s), fixation (2s), instruction (2s), task (4s). The session incorporates task for all fingers of the right hand with 25 tasks per finger and a total of 250 trials.} \label{fig1}
\end{figure}

\begin{figure}[t!] 
\centering
\includegraphics[width=\textwidth]{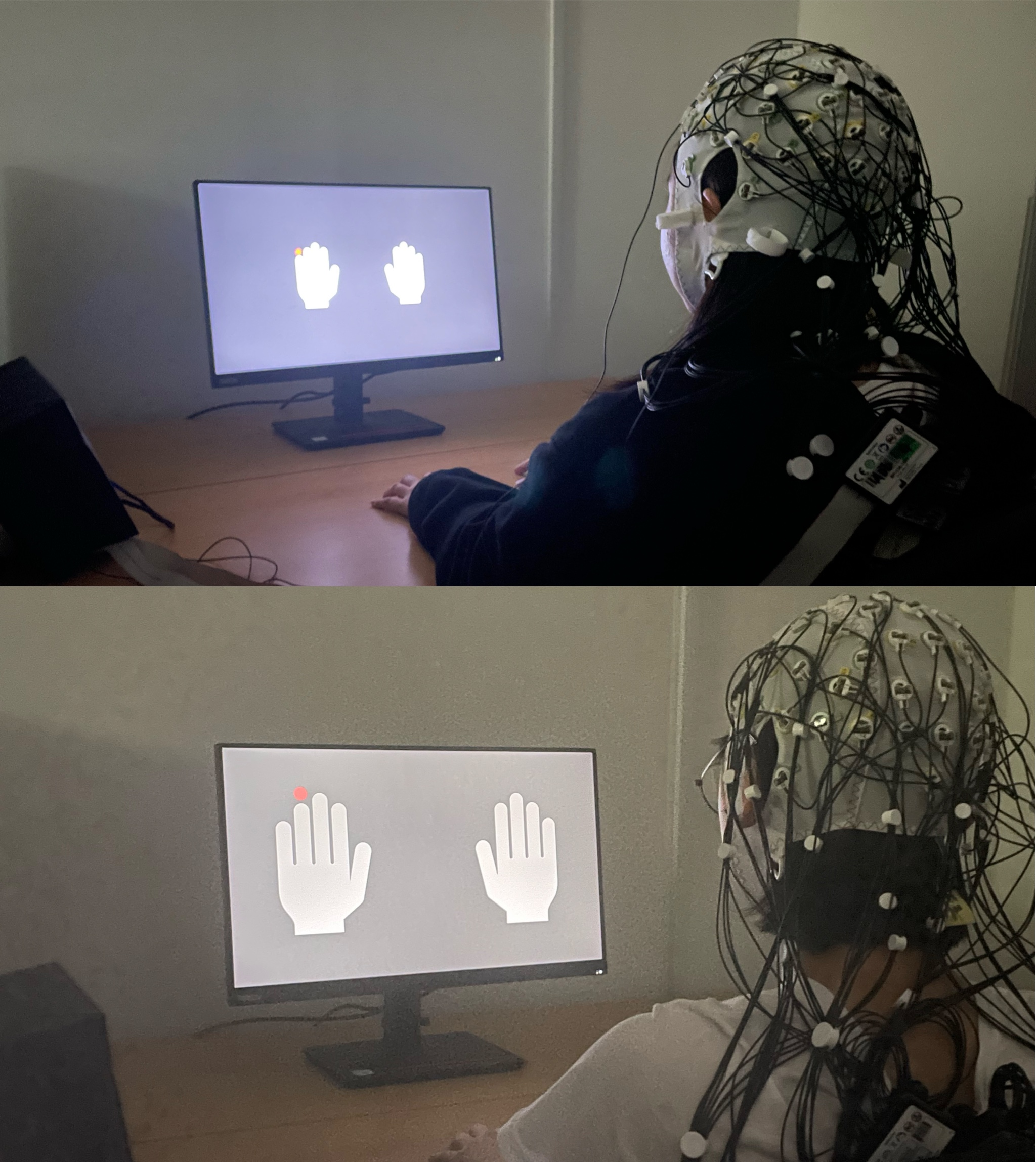}
\caption{Setting for acquiring EEG data during the experiment.} \label{fig2}
\end{figure}

Nine healthy participants, consisting of 4 males and 5 females, including 2 left-handed and 7 right-handed individuals, aged 20 to 25, participated in the experiments following a comprehensive briefing on the experimental paradigm and protocol. Fig. \ref{fig1} shows the overall experimental protocol representing the trial. This trial process is repeated 250 times. Fig. \ref{fig2} illustrates the actual experimental setup. 
Subjects were positioned in a room designed to minimize external noise and with subdued lighting conditions. They were directed to close their eyes and maintain a state of relaxation for a duration of 10 minutes, during which EEG data was collected. Before the initiation of the experiment, participants were directed to maintain a state of general well-being and obtain a sufficient duration of sleep, exceeding approximately 8 hours. Subsequently, written informed consent was obtained from each participant following the guidelines of the Declaration of Helsinki. The Institutional Review Board of the Chungbuk National University approved all experimental procedures under CBNU-202310-HRHR-0228.

We organized the experiment into ten classes, which covered the fingers of both the left and right hand. Within each session, the participants performed 25 tasks per finger. The experiment comprised two sessions: one that involved physical tapping of the finger and the other that involved the simulation of finger movements without physical execution. In this research, we employed only the data corresponding to the five classes (all fingers of the right hand) and the MI session.

\subsection{EEG Preprocessing}

EEG signals were recorded using a 64 channel EEG cap and BrainVision Recorder software (BrainProduct GMbH, Germany), operating at a sampling rate of 1000 Hz. The impedance for all channels was maintained below 10k$\Omega$. Raw EEG signals were processed in MATLAB R2022b(Mathworks Inc., USA) using the EEGLAB toolbox. A 60 Hz Notch filter was applied to reduce the noise of the power supply. Additionally, the raw EEG signals were downsampled to 250 Hz. We selected 24 channels (F3, F1, Fz, F2, F4, FC3, FC1, FC2, FC4, C3, C1, Cz, C2, C4, CP3, CP1, CPz, CP2, CP4, P3, P1, Pz, P2, and P4) considering their proximity to the motor cortex \cite{ref_article13,add_13_1,add_13_2}.

\subsection{Model Structure}

EEGNet is a convolutional neural network that uses depthwise convolution and separable convolution layers to extract spatial features from EEG signals. To enhance the classification performance of fine MI, we designed the network to extract higher-dimensional temporal features along with spatial features. This approach is derived from the main concept of DeepConvNet, which is to extract a diverse set of features without being limited to specific types. We derived inspiration for the design of FingerNet from these two baseline models. FingerNet, designed by incorporating features from EEGNet and DeepConvNet, is advantageous for fine MI classification by extracting spatial and temporal high-dimensional features from EEG signals. FingerNet extracts features through the same depthwise and separable convolution layers as EEGNet, followed by three pairs of convolution and pooling layers similar to DeepConvNet, enabling the extraction of higher-dimensional features. To reduce the complexity of multiple layers in FingerNet, we incorporated max norm and dropout techniques. Table I represents a structural comparison of EEGNet, DeepConvNet, and FingerNet. All three models employed identical activation functions, optimizers, and loss functions.

\begin{table}[hbt!]
\caption{Structural comparison of FingerNet and baseline networks: EEGNet, DeepConvNet}
\label{tab:table1}%
\renewcommand{\arraystretch}{1.2}
\resizebox{\linewidth}{!}{
\begin{tabular}{llll}
\hline
              & EEGNet                & DeepConvNet           & FingerNet             \\ \hline
              & Conv2D                & Conv2D                & Conv2D                \\
              & DepthwiseConv2D       & Conv2D                & DepthwiseConv2D       \\
              & AvgPool2D             & MaxPool2D             & AvgPool2D             \\
              & SeparableConv2D       & Conv2D                & SeparableConv2D       \\
              & AvgPool2D             & MaxPool2D             & AvgPool2D             \\
              & Fully-Connected Layer & Conv2D                & Conv2D                \\
Layers        & Softmax               & MaxPool2D             & AvgPool2D             \\
              &                       & Conv2D                & Conv2D                \\
              &                       & MaxPool2D             & AvgPool2D             \\
              &                       & Fully-Connected Layer & Conv2D                \\
              &                       & Softmax               & AvgPool2D             \\
              &                       &                       & Fully-Connected Layer \\
              &                       &                       & Softmax               \\ \hline
Activation    & ELU                   & ELU                   & ELU                   \\
Optimizer     & Adam                  & Adam                  & Adam                  \\
Loss function & Cross-entropy         & Cross-entropy         & Cross-entropy         \\ \hline
\end{tabular}}
\end{table}

\subsection{Loss Function}
The cross-entropy stands as a primary loss function in deep learning training, used to calculate the disparity between the actual and predicted values. It serves to evaluate the accuracy of the model's predictions and, in turn, guides the learning process. Cross-entropy places emphasis on penalizing erroneous predictions to a greater extent than on accurate predictions. This approach consequently presents a drawback in which the learning process is more prominent for majority classes, leading to a reduction in the amount of training dedicated to minority classes. For this constraint, the weighted cross-entropy is implemented, using the quantity of data associated with each class as a weighting factor to adjust the loss ratio. Derived from this perspective, we employed weighted cross-entropy by adjusting weights to prevent biased predictions. We induced the resolution of biased predictions by increasing the loss value when predictions for specific classes (thumb and index) fail through the utilization of weighted cross-entropy. This approach aims to address biased predictions by assigning higher penalties for wrong predictions associated with particular classes. The conventional weighted cross-entropy for addressing class imbalance adjusts weights based on the quantity of data, whereas weighted cross-entropy designed to prevent biased predictions, there was no established metric for setting weights. Therefore, we initially set the weights for each class to 1.0 and proceeded to heuristically adjust the weights by incrementally reducing the weight of the predominantly predicted class by 0.05 in order to mitigate biased predictions.

\begin{equation}
CE      = -\log(p_i, q_i) 
\end{equation}
\begin{equation}
WCE     = -\alpha_i \ast \log(p_i, q_i) 
\end{equation}
\begin{equation}
BWCE    = -\varpi_i \ast \log(p_i, q_i) 
\end{equation}

The formulations for cross-entropy, conventional weighted cross-entropy, and weighted cross-entropy designed to mitigate bias predictions are denoted as (1), (2), and (3), respectively. The symbols $q$ and $p$ represent the actual and predicted values, respectively. Where $\alpha$ corresponds to the reciprocal of the number of class data and $w$ denotes the heuristic weight values set in this study.

\begin{table}[ht]
\scriptsize
\caption{Performance comparison of FingerNet and baseline networks: EEGNet, DeepConvNet}
\label{tab:table2}
\renewcommand{\arraystretch}{1.2}
\centering
\resizebox{0.7\columnwidth}{!}{
\begin{tabular}{cccc}
\hline
          & EEGNet & DeepConvNet & FingerNet \\ \hline
Subject 1 & 0.2880 & 0.2960      & 0.3920   \\
Subject 2 & 0.2480 & 0.2800      & 0.2800   \\
Subject 3 & 0.2240 & 0.2400      & 0.2400   \\
Subject 4 & 0.2160 & 0.2240      & 0.2880   \\
Subject 5 & 0.2720 & 0.2320      & 0.3440   \\
Subject 6 & 0.2320 & 0.2480      & 0.3280   \\
Subject 7 & 0.2480 & 0.2480      & 0.2960   \\
Subject 8 & 0.2480 & 0.2800      & 0.3200   \\
Subject 9 & 0.2400 & 0.2320      & 0.2560   \\ \hline
Average &
  \begin{tabular}[c]{@{}c@{}}0.2196\\ ($\pm$0.0225)\end{tabular} &
  \begin{tabular}[c]{@{}c@{}}0.2533\\ ($\pm$0.0256)\end{tabular} &
  \begin{tabular}[c]{@{}c@{}}\textbf{0.3049}\\ ($\pm$0.0481)\end{tabular} \\ \hline
\end{tabular}}
\end{table}

\section{Results and Discussion}

\begin{figure}[t!]
\centering
\includegraphics[width=\textwidth]{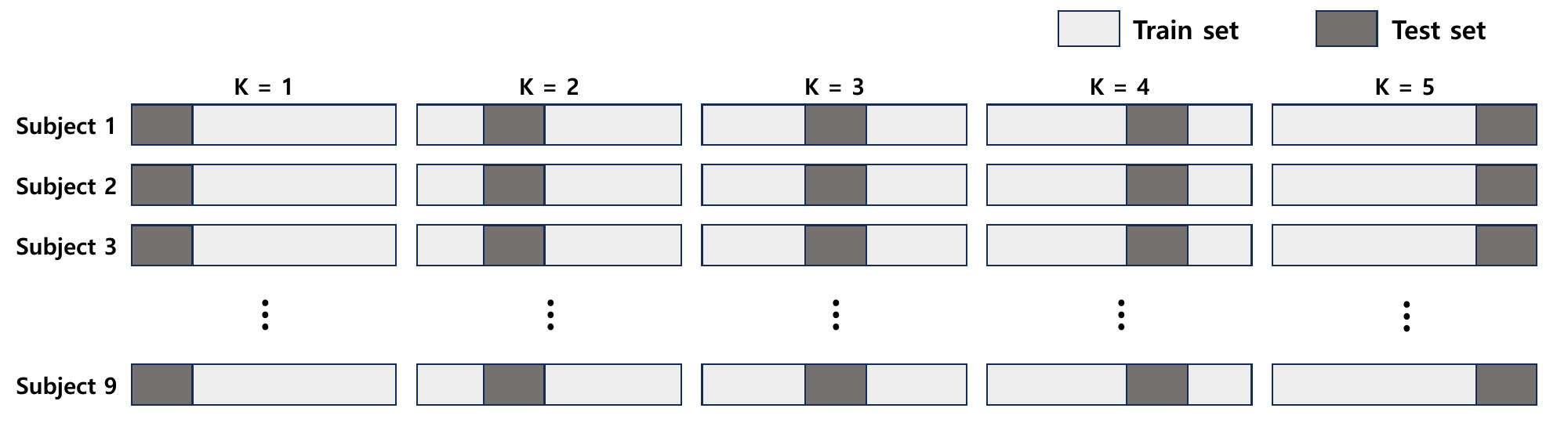}
\caption{Performance comparison of weighted cross-entropy using each set of weights.} \label{fig3}
\end{figure}

Initially, we compared the performance of EEGNet, DeepConvNet, and FingerNet. Table 2 presents a comparison of the performance of these three networks. To avoid the issues of under- and over-fitting, we applied 5-fold cross-validation. Fig. \ref{fig3} is a methodology employed for conducting cross-validation. We conducted cross-validation by setting aside a test set for subjects 1 to 9, utilizing an 8:2 ratio, allowing for the incorporation of the entire dataset as a test set. Experimental results indicate that FingerNet outperformed DeepConvNet and EEGNet. In Table 2, the average accuracy for FingerNet was 0.3049, whereas EEGNet and DeepConvNet exhibited lower accuracies of 0.2196 and 0.2533, respectively. In this result, the predominant performance of DeepConvNet over EEGNet suggests that temporal features are more effective than utilizing spatial features in fine MI classification. For these reasons, FingerNet, which adds convolution and pooling layers to extract high-dimensional temporal features, achieved the highest performance. In particular, in the case of Subject 1, the performance using FingerNet showed a performance improvement of approximately 0.1 or more compared to the use of other baseline networks. However, for Subject 3, the performance of FingerNet was not effective. The performance is comparable to the chance level, it is inferred that for Subject 3, none of the three networks successfully learned the task. Furthermore, for Subject 5, the right hand MI performance was expected to be low considering that the subject was left-handed, but it showed the second highest value among all subjects. Due to these factors, our proposed FingerNet demonstrated higher MI performance regardless of the dominant hand, outperforming baseline networks. We validated whether these performance differences are statistically significant through statistical testing. The predominance of FingerNet compared to baseline networks was substantiated by statistical validation carried out, demonstrating statistical significance differences ($p \leq $0.01).

\begin{figure}[t!]
\centering
\includegraphics[width=0.9\textwidth]{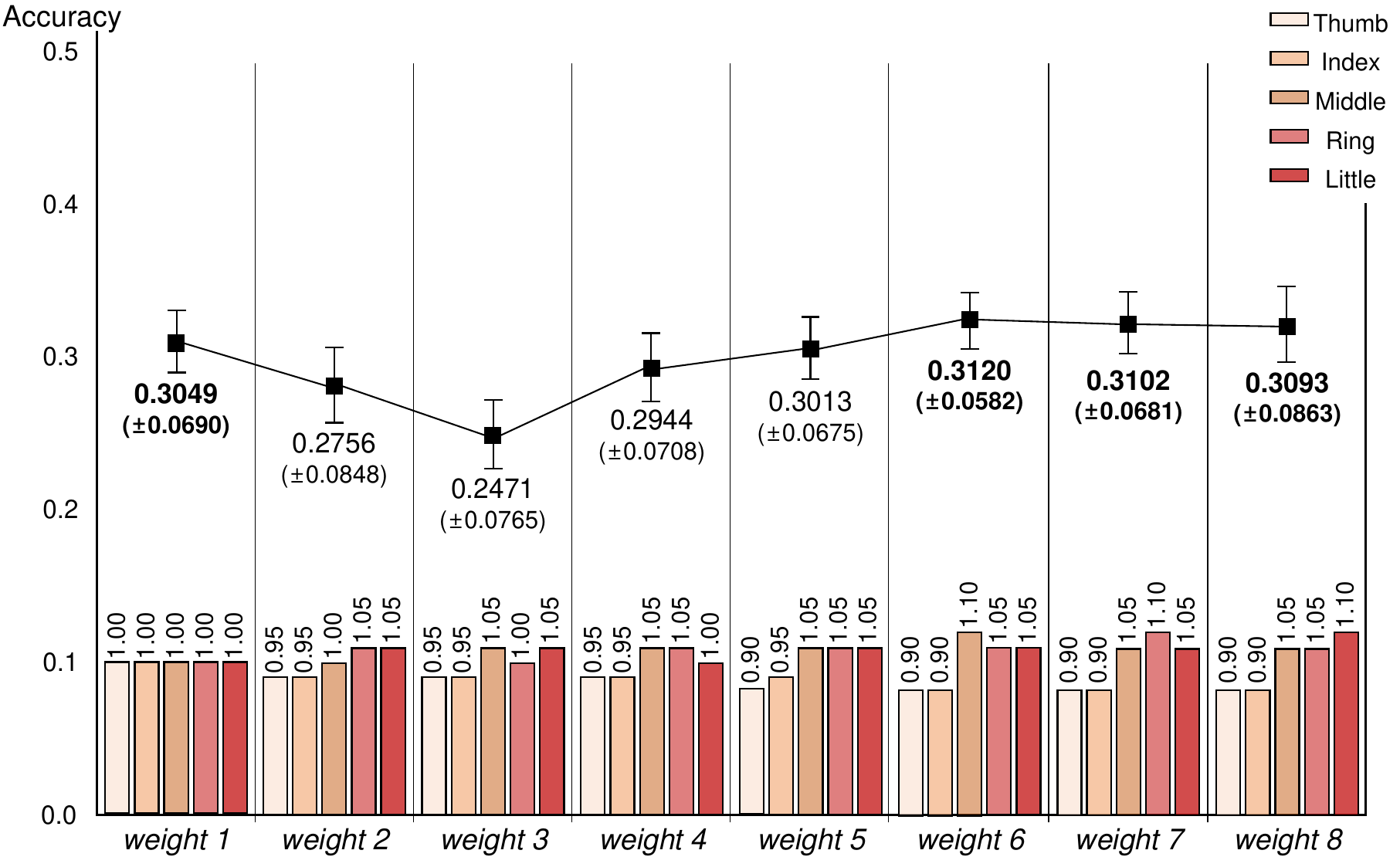}
\caption{Performance comparison of weighted cross-entropy using each set of weights.} \label{fig4}
\end{figure}

However, in FingerNet, upon closer examination of the predictions, certain issues were identified. FingerNet focused its predictions on the thumb and index classes. To address these issues, we employed weighted cross-entropy, which assigns weights to the conventional cross-entropy. These biased predictions suggest that FingerNet could not accurately capture features for other classes. Due to the considerable cost and time involved in obtaining additional EEG data, we have designed a method to modify the loss function, enabling FingerNet to make more balanced predictions. We heuristically adjusted the weights for each class in increments of 0.05, employing a strategy that reduces the weights for classes with relatively higher prediction frequencies such as thumb and index, and increases the weights for middle, ring, and little. Fig. \ref{fig4}, the performance for each set of weights is illustrated, showing that the results of training with varying weights indicate that FingerNet surpassed its performance with the conventional cross-entropy when using $weight$ 6, 7, and 8. Weights 6, 7, and 8 exhibited dominanat performance with values of 0.3120, 0.3102, and 0.3093, respectively, surpassing the conventional cross-entropy loss function.

\begin{figure}[t!]
\centering
\includegraphics[width=0.9\textwidth]{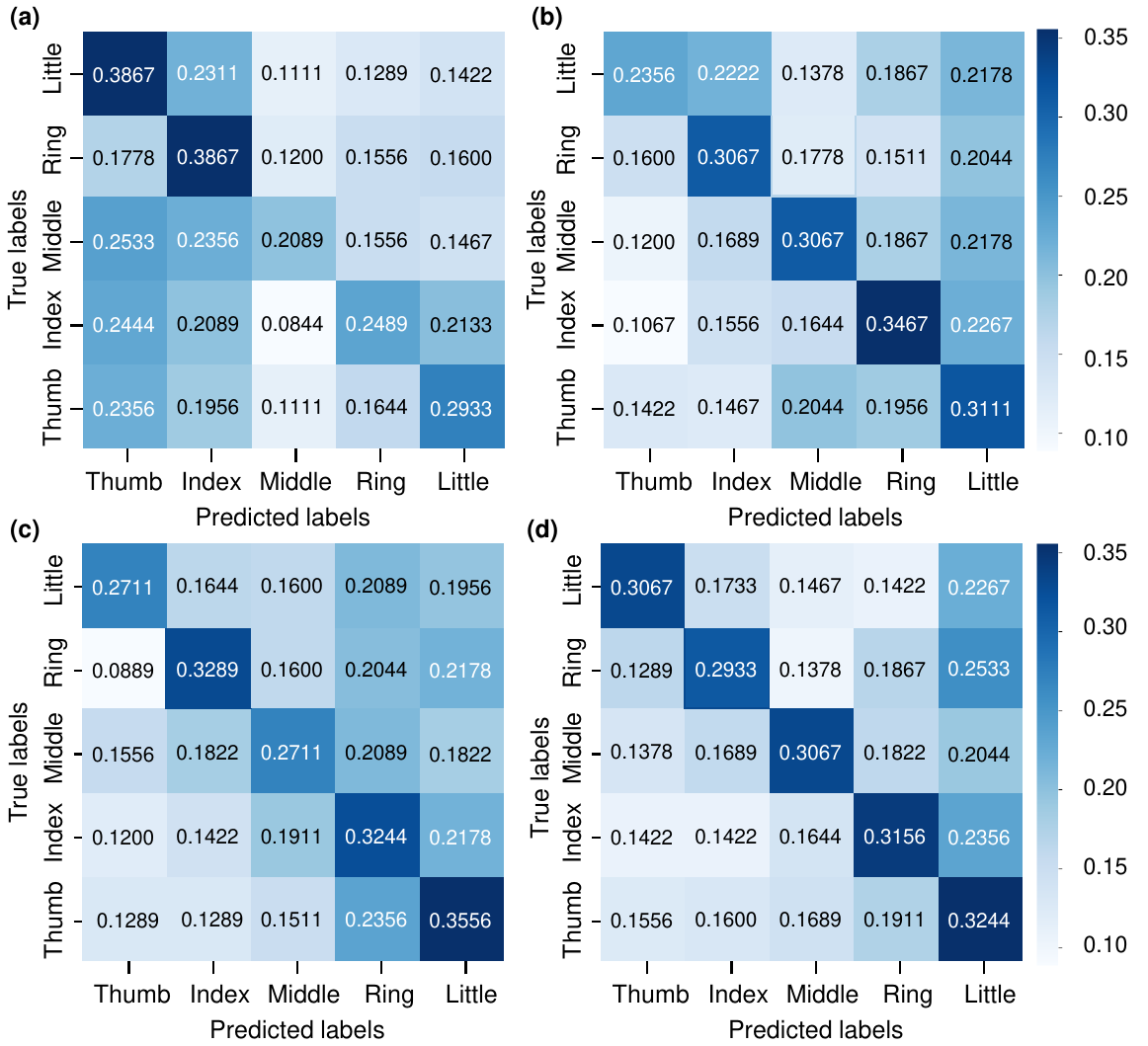}
\caption{Confusion matrix using weighted cross-entropy with each set of weights: (a) used conventional cross-entropy, while (b), (c), and (d) employed weighted cross-entropy with the inclusion of weights 6, 7, and 8.} \label{fig5}
\end{figure}

To figure out the effectiveness of adjusting these weights in mitigating biased predictions, we evaluated the confusion matrices for weight 1 (utilizing the conventional cross entropy loss function) and weights 6, 7, and 8. Fig. \ref{fig5} is a confusion matrix utilizing each set of varying weights. Fig. \ref{fig5} (a), biased predictions were observed toward the thumb and index classes. However, adjusting the weights showed balanced predictions. Fig. \ref{fig5} (b), it is evident that adding weights to the middle class, which was initially predicted to be the least, also resulted in a more balanced prediction for the middle class. In particular, it was observed that the performance of the middle and ring classes in Fig. \ref{fig5} (a) was 0.2089 and 0.2489, respectively, whereas in Fig. \ref{fig5} (b) after weight adjustment, it increased to 0.3067 and 0.3467. In contrast, in the cases of Figs. \ref{fig5} (c) and (d), the increase in weights for the ring and little classes resulted in improved accuracy. Especially, we confirmed predictions of 0.2 or more for all instances of the little class. However, paradoxically, the predictions were predominantly concentrated on the ring and little classes. This suggests that excessive adjustment of weights led to biased predictions towards another class. As indicated by these results, modifying the weights can mitigate the model's biased predictions, but the biased prediction may appear in the opposite direction. By addressing these aspects and optimizing weights or heuristically adjusting them for a model, it may be possible to alleviate the model's biased predictions. 

%
%

\section{Conclusion and Future Works}
In this paper, we proposed FingerNet which is a network specialized for fine MI classification using temporal features. FingerNet surfaced predominant performance in same hand classification when compared to other baseline models: EEGNet and DeepConvNet. Although its current performance is limited by relatively low performance, the model suggested the potential for classifying MI in localized regions by utilizing high-dimensional temporal features along with spatial features. Furthermore, in this study, the utilization of a weighted cross-entropy loss function, conventionally employed to address class imbalance, was suggested as a means to mitigate the model's biased predictions. This approach helps achieve balanced predictions by adjusting the loss function in multi-class classification, where biased predictions occur. This implies that modifying the loss function to address biased predictions in scenarios where data acquisition, such as EEG data, is cost-intensive, could be a beneficial approach. In future work, we will improve the accuracy of FingerNet. As mentioned earlier, it is important to design a more efficient network structure. Moreover, in this experiment, there was an imbalance in the ratio between left-handed and right-handed subjects. To address the inadequacy in assessing MI abilities based on the dominant hand, we plan to collect a larger and more balanced dataset in future work. In conclusion, through such efforts to enhance performance, we believe that effective execution of motor imagery for not only fine MI, but also gross MI can be achieved.
\\

\begin{credits}
\textit{*Corresponding author: Ji-Hoon Jeong}

\end{credits}
%
%
%
%

\end{document}